\documentstyle[11pt,newpasp,twoside,psfig]{article}
\index{instructions}
\index{guidelines}

\index{radio}
\index{X-rays}
\index{ROSAT}
\index{Chandra}
\index{XMM-Newton}

\index{shocks}
\index{shock acceleration}
\index{synchrotron emission}
\index{neutron stars!statistics}
\index{central compact objects}
\index{central compact objects!associations with supernova remnants}

\index{pulsars}
\index{pulsars!surveys}
\index{pulsars!statistics}
\index{pulsars!ages}
\index{pulsars!winds}
\index{pulsars!spectra}
\index{pulsars!spin-down}
\index{pulsars!emission models}
\index{pulsars!associations with supernova remnants}
\index{pulsars!interactions with supernova remnants}

\index{pulsars!J0537--6910}
\index{pulsars!J0534+2200}
\index{pulsars!J0540-6919}
\index{pulsars!J0205+6449}
\index{pulsars!J2229+6114}
\index{pulsars!J1513--5908}
\index{pulsars!J1617--5055}
\index{pulsars!J1124--5916}
\index{pulsars!J1930+1852}
\index{pulsars!J1420--6048}
\index{pulsars!J1846--0258}
\index{pulsars!J0835--4510}
\index{pulsars!J1811--1925}
\index{pulsars!J1112--6103}
\index{pulsars!J1952+3252}
\index{pulsars!J1702--4429}
\index{pulsars!J2021+3651}
\index{pulsars!J1524--5706}
\index{pulsars!J1913+1011}
\index{pulsars!J1836--1334}
\index{pulsars!J1801--2451}
\index{pulsars!J1016--5857}
\index{pulsars!J1103--6107}
\index{pulsars!J1119--6127}
\index{pulsars!J1803--2137}
\index{pulsars!J1048--5832}
\index{pulsars!J1837--0604}
\index{pulsars!J0940--5428}
\index{pulsars!J1856+0113}

\index{pulsar wind nebulae!wisps}
\index{pulsar wind nebulae!spectra}
\index{pulsar wind nebulae!spectral index variations}
\index{pulsar wind nebulae!time evolution}
\index{pulsar wind nebulae!interactions with supernova remnants}

\index{pulsar wind nebulae!N157B}
\index{pulsar wind nebulae!Crab Nebula}
\index{pulsar wind nebulae!B0540--69}
\index{pulsar wind nebulae!3C58}
\index{pulsar wind nebulae!G106.6+2.9}
\index{pulsar wind nebulae!MSH~15--52}
\index{pulsar wind nebulae!G292.0+1.8}
\index{pulsar wind nebulae!G54.1+0.3}
\index{pulsar wind nebulae!Kookaburra}
\index{pulsar wind nebulae!Kes~75}
\index{pulsar wind nebulae!Vela}
\index{pulsar wind nebulae!G11.2--0.3}
\index{pulsar wind nebulae!CTB~80}

\index{supernova remnants}
\index{supernova remnants!surveys}
\index{supernova remnants!statistics}
\index{supernova remnants!shell}
\index{supernova remnants!morphology}
\index{supernova remnants!extragalactic}
\index{supernova remnants!associations with pulsars}
\index{supernova remnants!interactions with pulsar wind nebulae}

\index{supernova remnants!N157B}
\index{supernova remnants!Crab}
\index{supernova remnants!Crab Nebula}
\index{supernova remnants!B0540--69}
\index{supernova remnants!3C58}
\index{supernova remnants!G106.6+2.9}
\index{supernova remnants!MSH~15--52}
\index{supernova remnants!G292.0+1.8}
\index{supernova remnants!G54.1+0.3}
\index{supernova remnants!Kookaburra}
\index{supernova remnants!Kes~75}
\index{supernova remnants!Vela}
\index{supernova remnants!G11.2--0.3}
\index{supernova remnants!CTB~80}
\index{supernova remnants!G292.2+0.5}     
\index{supernova remnants!W44}

\def\cxo{{\em Chandra}}

\def \s{\phantom}
\def\simgt{\lower.5ex\hbox{$\; \buildrel > \over \sim \;$}}
\def\simlt{\lower.5ex\hbox{$\; \buildrel < \over \sim \;$}}

\def\edcomment#1{\iffalse\marginpar{\raggedright\sl#1\/}\else\relax\fi}
\reversemarginpar

\begin{document}
\title{A Spin-down Power Threshold for Pulsar Wind Nebula Generation?}
\author{E. V. Gotthelf}
\affil{Columbia Astrophysics Laboratory, Columbia University, 505 West 120$^{th}$ Street, New York, NY 10027, USA}

\begin{abstract}
A systematic X-ray survey of the most energetic rotation-powered
pulsars known, based on spin-down energy loss rate, $\dot E =
I\omega\dot\omega$, shows that all energetic pulsars with $\dot E >
\dot E_{c} \approx 3.4 \times\ 10^{36}$ erg s$^{-1}$ are X-ray bright,
manifest a distinct pulsar wind nebula (PWN), and are associated with
a supernova event, either historically or via a thermal remnant, with
over half residing in shell-like supernova remnants.  Below $\dot
E_{c}$, the $2-10$ keV PWN flux ratio $F_{PWN}/F_{PSR}$ decreases by
an order-of-magnitude.  This threshold is predicted by the lower limit
on the spectral slope $\Gamma_{min} \approx 0.5$ observed for
rotation-powered pulsars (Gotthelf 2003). The apparent lack of bright
pulsar nebulae below a critical $\dot E$ suggests a change in the
particle injection spectrum and serves as a constraint on emission
models for rotation-powered pulsars.  Neither a young age nor a high
density environment is found to be a sufficient condition for
generating a PWN, as often suggested, instead the spin-down energy
loss rate is likely the key parameter in determining the evolution of
a rotation-powered pulsar.

\end{abstract}

\vspace {-0.25cm}
\section{A \cxo\ Study of the Most  Energetic Pulsars}

Table~1 presents the 28 most energetic pulsars from the complete
pulsar catalog of Manchester (2003), ordered by spin-down power ($\dot
E = I\omega\dot\omega$, where $I$ is the neutron star moment of
inertia and $\omega$ is its angular velocity). These include all known
pulsars detected in both the radio and X-ray energy bands with a
spin-down power above $\dot E = 10^{36}$ erg s$^{-1}$ (but excludes
the one millisecond pulsar in this range).  Of the full list, 25 out
of 28 sample objects are radio pulsars, 21 are X-ray pulsars, of which
only 3 are detected in X-rays alone. So far, 5 radio pulsars have no
known follow-up yet in any waveband. For each pulsar with available
\cxo\ ACIS X-ray data, and for its PWN, we measured the unabsorbed
flux in the $2-10$ keV band using the method described in Gotthelf
(2003). Herein, we compared these fluxes with the spin-down energy
loss rate and present the flux ratio $F_{PWN}/F_{PSR}$, where
$F_{PSR}$ is the sum of the pulsed and unpulsed pulsar emission.

All of the top 13 pulsars in Table~1 have been observed in X-rays;
this includes the 9 brightest X-ray PWN used in the initial study of
Gotthelf (2003). When ordered by $\dot E$ it is apparent that all
energetic pulsars with $\dot E_{c} \simgt 3.4 \times\ 10^{36}$ erg
s$^{-1}$ are X-ray bright, show a resolved PWN, and are associated
with evidence of a supernova event. The jury is still out on
PSR~J1617$-$5055, which is highly 
%\noindent 
absorbed and was observed with \cxo\ too far
off-axis to resolve a nebula, and on J1112$-$6102, for
which no follow-up X-ray observation currently exist.

%\clearpage

\begin{table}[ht]
%\footnotesize
\small
\begin{tabular}{llcccccl}
\multicolumn{7}{c}{\bf Table 1: Pulsars Ordered by Spin-down Power$^a$}\\ [0.1in]
\tableline\tableline
 Pulsar      & Remnant       & $\dot E^a$        & Dist$^b$   & $\epsilon^c =$  & $F_{PWN}/$ & Code$^d$ \\
             &               &$\times 10^{36}$   &            & $L_X/\dot E$ & $F_{PSR}$  &          \\
             &               &(erg/s)            & (kpc)      &                  &            &          \\      
\tableline
J0537$-$6910 & N157B         &      481.6        & 49         & 0.003            & 15    & {\tt s-x}\\
J0534$+$2200 & Crab (SN1054) &      440.6        & 2.0        & 0.03             & 30    & {\tt srx}\\
J0540$-$6919 & SNR~0540$-$69 &      146.5        & 49         & 0.05             & 4     & {\tt srx}\\
J0205$+$6449 & 3C58 (SN1181) &   \s\ 27.0        & 3.2        & 0.0004           & 60    & {\tt srx}\\
J2229$+$6114 & G106.6$+$2.9  &   \s\ 22.5        & 12         & 0.001            & 9     & {\tt -rx}\\
J1513$-$5908 & MSH 15$-$52   &   \s\ 17.7        & 5.0        & 0.01             & 5     & {\tt srx}\\
J1617$-$5055 &               &   \s\ 16.2        & 6.5        & 0.001            & \dots & {\tt -rx}\\
J1124$-$5916 & G292.0$+$1.8  &   \s\ 11.9        & 5.4        & 0.0002           & 10    & {\tt -rx}\\
J1930$+$1852 & G54.1$+$0.3   &   \s\ 11.6        & 5          & 0.002            & 5     & {\tt srx}\\
J1420$-$6048 & Kookaburra    &   \s\ 10.4        & 7.7        & 0.004            & 10    & {\tt -rx}\\
J1846$-$0258 & Kes~75        &\s\ \s\ 8.3        & 19         & 0.15             & 23    & {\tt s-x}\\
J0835$-$4510 & Vela~SNR      &\s\ \s\ 6.9        & 0.3        & 0.0001           & 9     & {\tt srx}\\
J1811$-$1926 & G11.2$-$0.3 (SN386?) &\s\ \s\ 6.4 & 5          & 0.006            & 9     & {\tt s-x}\\
J1112$-$6103 &               &\s\ \s\ 4.5        & \dots      & \dots 	         & \dots & {\tt -r-}\\
J1952$+$3252 & CTB~80        &\s\ \s\ 3.7        & 2.5        & 0.0005           & 1.1   & {\tt -rx}\\
\tableline							      	      
J1709$-$4429 & G343.1$-$2.3? &\s\ \s\ 3.4        & 2.5        & 0.0001           & 3.5   & {\tt -rx}\\
J2021$+$3651 &               &\s\ \s\ 3.4        & 10         & \dots 	         & \dots & {\tt -r?}\\
J1524$-$5625 &               &\s\ \s\ 3.2        & 3.8        & \dots 	         & \dots & {\tt -r?}\\
J1913$+$1011 &               &\s\ \s\ 2.9        & 4.5        & \dots 	         & \dots & {\tt -r?}\\
J1826$-$1334 &               &\s\ \s\ 2.9        & 4.1        & 0.0008           & 2.3   & {\tt -rx}\\
J1801$-$2451 &               &\s\ \s\ 2.6        & 4.6        & 0.0008           & 0.1   & {\tt -rx}\\
J1016$-$5857 &               &\s\ \s\ 2.6        & 9.3        & \dots 	         & \dots & {\tt -rx}\\
J1105$-$6107 &               &\s\ \s\ 2.5        & 7.1        & \dots            & \dots & {\tt -r-}\\
J1119$-$6127 & G292.2$-$0.5 (radio) &\s\ \s\ 2.3 & 4          & 0.00005          & 0.2   & {\tt -rx}\\
J1803$-$2137 &               &\s\ \s\ 2.2        & 4.0        & \dots 	         & \dots & {\tt -rx}\\
J1048$-$5832 &               &\s\ \s\ 2.0        & 3.0        & \dots 	         & \dots & {\tt -rx}\\
J1837$-$0604 &               &\s\ \s\ 2.0        & 6.2        & \dots 	         & \dots & {\tt -r?}\\
J0940$-$5428 &               &\s\ \s\ 1.9        & 4.3        & \dots            & \dots & {\tt -r?}\\ 
\tableline
\end{tabular}								  
\\
\footnotesize
$^a${Table rank-ordered by spin-down power $\dot E = I\omega\dot\omega$, were $I \equiv 10^{45}$ gm cm$^{-2}$.} \\
$^b${Best estimate of the pulsar distance ($d$, in kpc) from the literature.} \\
$^c${Efficiency, $\epsilon$, the ratio of pulsar luminosity ($L_X \equiv F_X/4\pi d^2 = L_{PWN}+L_{PSR}$) in the $2-10$ keV band measured following the procedure of Gotthelf (2003)
and the spin-down power.} \\ 
$^d${Code: {\tt s}=\cxo\ PWN survey object (Gotthelf 2003); {\tt r}=Radio source; {\tt x}=X-ray source.}\\
%\vskip -0.5cm
\end{table}

In contrast, pulsars whose spin-down energy loss rate falls below
$\dot E_{c}$ lack both a bright nebula and a supernova association in
the X-ray energy regime. For several of these objects \cxo\
observations detect weak nebulously. Diffuse X-ray emission is found
around PSR~J1709$-$4420 (Gotthelf et al. 2002) and tentatively
indicted for PSR~J2021$+$3651, a newly discovered pulsar with a
similar $\dot E$ (Roberts, this proceedings).  An extremely faint
X-ray ``tail'' is found trailing the ``Duck'' radio pulsar
PSR~J1801$-$2451, but this is interpreted as a ram-pressured confined
cometary wind (Kaspi et al. 2001). The \cxo\ observation of
PSR~J1826$-$1334 confirms a faint PWN, barely resolved with the ROSAT
HRI (Finley et al. 1996). Finally, arcsecond localization of
J1105$-$6107, previously associate with X-ray emission (Gotthelf \&
Kaspi 1998), shows that the X-rays originate from an unrelated nearby
source. Of the remaining pulsars below the $\dot E_c$ line in Table 1,
none have arguably a definitive PWN or SNR association in X-rays.

Evidently all pulsars with $\dot E > \dot E_c$ display bright PWNe
while for the less energetic pulsars the nebula emission is vestigial,
at best, when resolved from the background. This fact is quantified by
the flux ratio $F_{PWN}/F_{PSR}$ given in Table 1 which shows that the
PWNe of the less energetic pulsars are genuinely sub-luminous relative
to their PSR flux.  This comparison is best done statistically since
the distance estimates are mostly uncertain (factor $\sim 2$). Above
$\dot E_c$, the mean flux ratio for these pulsars is of order $\sim
14$, while the less energetic pulsars have a ratio of order $\sim
1.5$. This factor of ten change in efficiency in the X-ray band cannot
be explained as a distance bias, i.e., the bright PWNe are
systematically closer, as the range of distances overlap between the
less and more energetic pulsars (see Table 1). Deeper observations of
the faint PWNe are needed to search for extended emission yet missed.

%-----------------------------Figure Start--------------------------------
\begin{figure}[t]
\begin{center}
\centerline{
  \psfig{figure=gotthelfe_f1.ps,height=5.2cm,angle=0}
  \hspace{0.2cm}
  \psfig{figure=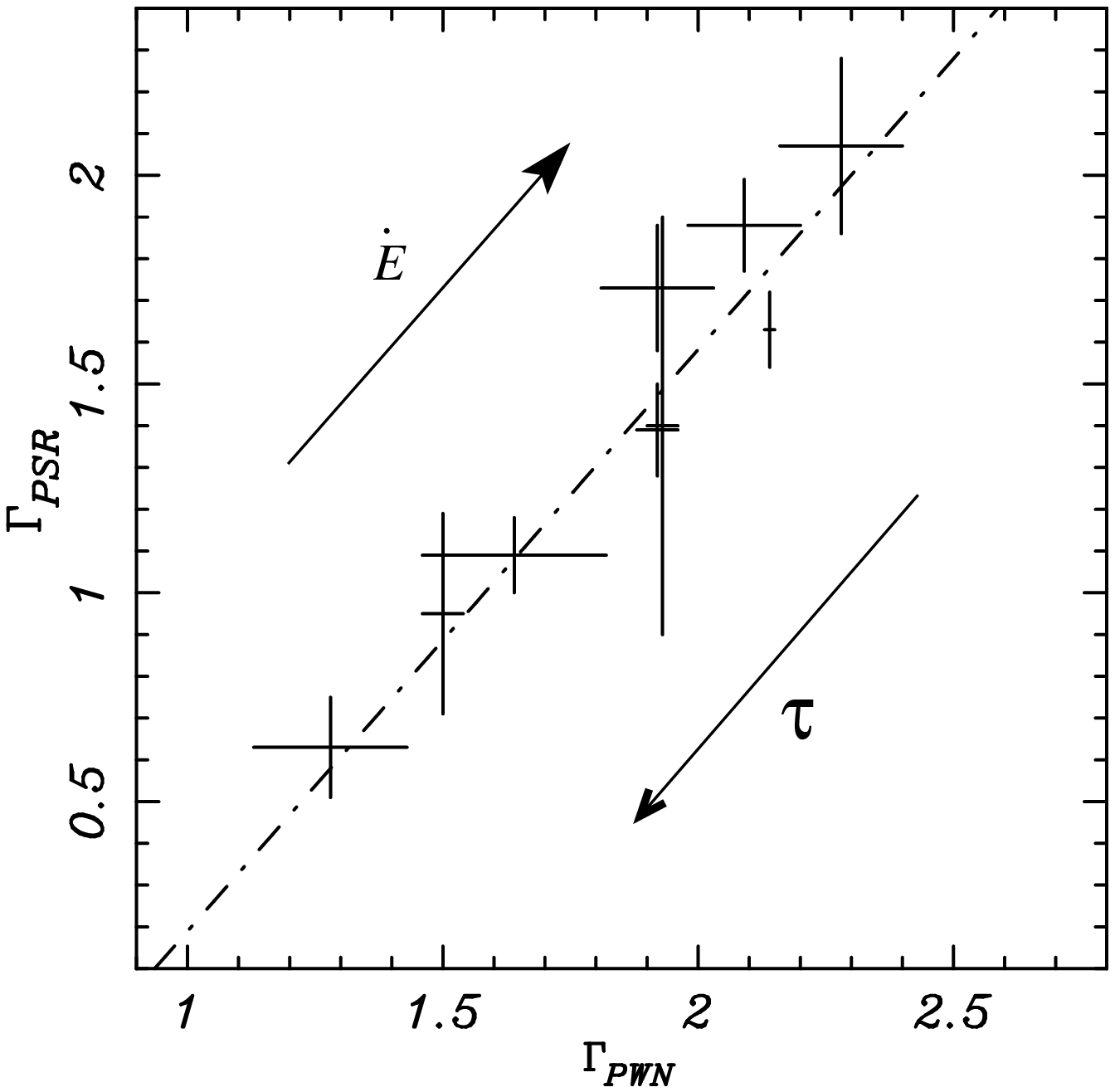,height=5.2cm,angle=0}
}
\vspace{-1.0cm}
\end{center}
\caption{
\footnotesize {{\it Left panel}: {A comparison between the
$2-10$ keV spectral slope of the nine brightest known pulsars
($\Gamma_{PSR}$) and the inverse square root of their spin-down power,
$\dot E_{40}^{-1/2}$, in units of $10^{40}$ erg s$^{-1}$. The
dashed-line indicates the best-fit model.}  {\it Right panel}:
{Relationship between the above pulsars' spectral slope ($\Gamma_{\rm
PSR}$) and that of their wind nebulae ($\Gamma_{\rm PWN}$), assuming a
simple power-law spectral model. The dashed line indicates the
best-fit. The physical origin of this relationship has yet to be determined. 
From Gotthelf (2003).  } }}
\label{fig1}
\vspace {-0.3cm}
\end{figure}

%-----------------------------Figure End----------------------------------

A possible explanation for a critical $\dot E_c$ is provided in
Gotthelf (2003), where the spectra of the most energetic pulsars are
shown to depend on their spin-down power $\dot E$ - the more energetic
the pulsar, the steeper its spectral slopes. This rule follows an
inverse square-root law, $\Gamma_{PSR} = \Gamma_{max} + \alpha \dot
E^{-1/2}$ with a minimum observed spectral slope of $\Gamma_{min} \approx
0.5$ (see Fig. 1). Most interestingly, $\Gamma_{min}$ corresponds to
$\dot E_{c} \approx 3.4 \times\ 10^{36}$ erg s$^{-1}$, right at the observed
threshold for bright PWNe. Since the spectral index likely reflects
the spectrum of the injected wind particles (Pacini \& Salvati 1973),
a critical phenomena in the acceleration process may be responsible
for the observed threshold, perhaps turning off the pulsar wind or the
PWN shock and allowing the nebula to fade with time and/or $\dot E$.
For this fossil PWN, the above $\Gamma\ \rm{vs}\ \dot E^{-1/2}$
relationship likely becomes invalid; some evidence for this is
provided by preliminary spectra of faint nebulae belonging to the less
energetic pulsars.

The basic result presented herein is also seen in the radio waveband
where only the most energetic rotation-powered pulsars are found to
display a radio PWN (Cohen et al 1983; Frail \& Scharringhausen 1997;
Gaensler et al. 2000). The $\dot E_c$ threshold is also found to be
%applicable at these wavelengths, as none of the similar pulsars with a
%spin-down power below the critical value display any radio emission
%associated with a PWN at all, despite a sensitive search at 1.4 GHz
applicable at these wavelengths, as none of the less energetic pulsars
display a radio PWN at all, despite a sensitive search at 1.4 GHz
around 27 pulsars with $1.2 \times 10^{32} < \dot E < 2.8 \times
10^{36}$~erg/s by Gaensler et al. (2000). Possible exceptions are
PSR~J0908$-$4913, a pulsar with weak ($F_{PWN}/F_{PSR} < 1/16$ @
1.2--2.2 GHz), barely resolved radio emission (Gaensler et al. 1998),
and PSR~J1856$+$0113 in SNR W44 with an apparent PWN.  The latter
object, however, is unusual and its exact nature 
requires further study.
%investigation.

Because the $\dot E$ of the pulsars in the survey herein are unlikely
to be correlated with the local density, this parameter is not a
key factor for producing a detectable radio PWN as often
claimed. Nor is a young age likely a sufficient condition for
generating a PWN, considering the examples of PSR~J1119$-$6127, a
young pulsar ($P / 2 \dot P = 1.6$~kyr) in the radio shell
G292.2$-$0.5 lacking a PWN (e.g. Crawford et al. 2001).

\vspace {-0.1cm}
\section{Conclusions}

\begin{itemize}

\item The spin-down energy loss rate is a key evolutionary parameter
for rotation-powered pulsars.

\item A threshold exist $\dot E_{c} \approx 3.4 \times\
10^{36}$ erg s$^{-1}$ below which the generation of a PWN is 
greatly reduced (in X-rays) and/or undetected (in radio).

\item A Crab-like pulsar is defined as a rotation-powered pulsars
with \hbox{$\dot E > \dot E_{c}$}.

\item A young age or a high local density environment is not a
sufficient condition for generating a PWN, as often suggested.

\end{itemize}

\acknowledgments
This research is funded by LTSA NAG~5-7935.


\begin{references}
%\footnotesize
\reference Cohen, N.~L., et al. 1983, ApJ, 264, 273
\reference Crawford, C., et al. 2001, ApJ, 554, 152
\reference Finley, J.~P., et al. 1996 ApJ, 466, 938
\reference Frail, D.~A. \& Scharringhausen B.~R. 1997, ApJ, 480, 364
\reference Gaensler, B.~M., et al. 1998, ApJ, 499, L69
\reference Gaensler, B.~M., et al. 2000, MNRAS, 318, 58
\reference Gotthelf, E.~V. \& Kaspi, V.~M. 1998, ApJ, 497, L29
\reference Gotthelf, E.~V. 2003 ApJ, 591, 361
\reference Gotthelf, E.~V., Halpern, J.~P. \& R. Dodson 2002, ApJ, 567, L125
\reference Manchester, R.~N. 2003 {\tt www.atnf.csiro.au/research/pulsar/catalogue/}
\reference Kennel, D.~F. \& Coroniti, F.~V. 1984a, 283, 694; 1984b, 283, 710
\reference Kaspi, V.~M., et al. 2001, ApJ, 562, L163
\reference Pacini, F. \& Salvati, M. 1973 ApJ, 186, 249; 1973 ApL, 15, 39

\end{references}
\end{document}